\documentclass[global,twocolumn,final]{svjour}
\usepackage{graphics}
\usepackage{amsmath}

\journalname{Applied Physics B}
\begin{document}
\title{An optical multimode fiber as pseudothermal light source}
\author{Thomas Mehringer\inst{1,2} \and Steffen Oppel\inst{1,2} \and Joachim von Zanthier\inst{1,2}
}                     
%
%
\institute{Institut f\"ur Information und Photonik, Universit\"at Erlangen-N\"urnberg,
	91058 Erlangen, Germany \and Erlangen Graduate School in Advanced Optical Technologies (SAOT), Universit\"at Erlangen-N\"urnberg,
	91052 Erlangen, Germany}
\date{Received: date / Revised version: date}
%
\maketitle
\begin{abstract}
We report on a novel pseudothermal light source based on laser light coupled into an optical multimode fiber. The setup is simple, of low cost, exhibits inherently high directional light emission and allows for a flexible arrangement. By measuring the photon statistics and spatial two point intensity correlations in the far field we show that the setup exhibits all characteristics of a gaussian random source. 
\end{abstract}
\section{Introduction}
\label{sec:intro}
In the seminal experiment by Hanbury Brown and Twiss \cite{hbt1956correlations,hbt1956sirius} thermal light sources (TLS) have been used to investigate photon correlations and statistical problems as the 
number of photons in a given time interval or
time delays between successive photon counts. The experiment is widely believed to mark the beginning of quantum optics \cite{Glauber2005}. However, since then single photon emitters or entangled photon states have been at the focus in quantum optics due to the belief that quantum optical phenomena require non-classical states. This changed at the beginning of the century when ghost imaging with pseudothermal light was experimentally demonstrated \cite{Boyd2002}. Many quantum optical phenomena using 
pseudothermal light fields have been observed since \cite{Scarcelli_1TLS_2004,Scarcelli_2TLS_2004,Gatti2004,Ferri2005,Agafonov2008,Chen2009,Zhou2010,Oppel2012,Oppel2014,Bhatti2016}, and quantum interferences have been even demonstrated with true thermal light \cite{Chen2009true_thermal_light,Kurtsiefer2014,Zhang2005,Karmakar2012,Liu2014}. 
Albeit displaying a lower visibility,
the signal-to-noise ratio of pseudothermal light fields exceeds typically those of non-classcial fields by orders of magnitude.

Pseudo-TLS have been realized in different ways. The most common method employs a single mode laser impinging on a rotating ground glass disk (RGD) \cite{Martienssen1964,Arecchi1969,Estes1971,Rousseau1971}.
In this case a large number of speckles is produced due to stochastically interfering waves scattered from the granular surface of the RGD which act like many independent pointlike subsources.
 By rotating the disk the intensity at a given point fluctuates similarly to the intensity of a true thermal field. The advantage of such pseudo-TLS is that the coherence time can be adjusted by simply varying the angular velocity of the disk. 
A different pseudo-TLS has been developed recently using a laser beam in combination with a digital micro-mirror device \cite{Boyd2016}. Here, random binary diffraction structures (having Kolmogorov statistics) are imprinted onto a laser field resulting in a pseudothermal field distribution.

In this paper we  present a new pseudo-TLS based on coupling laser light into an optical multimode fiber. 
The advantages of the fiber pseudo-TLS is its simplicity and low cost, a high directional emission of photons at the fiber output as well as a high flexibility, especially if the source arrangement is to be changed repeatedly.

Currently, there is a demand for pseudo-TLS as real thermal sources display coherence times on the order of $\propto 10^{-10}-10^{-12}$ s and radiate polychromatic. This requires a) very fast electronics to make interference effects visible and b) spectral filtering what reduces the count rate \cite{Chen2009true_thermal_light,Kurtsiefer2014}. To overcome these problems one of the established pseudo-TLS described above can be used or the novel fiber pseudo-TLS described in this paper.

The paper is organized as follows: In Sect. 2 we present a simple model explaining the main features of the fiber pseudo-TLS, e.g., the speckle pattern and the photon statistics observed in the far field. In Sect. 3 we show experimental results of two point intensity correlations for one and two fiber pseudo-TLS corroborating the anticipated outcome of the pseudo-TLS.
In Sect. 4 we present our conclusions.
\section{Generation of  pseudothermal light using multimode fibers}
\label{sec:fiber_as_TLS}
	
The standard description of a monochromatic TLS is as follows \cite{Loudon2000}: light is produced by a large number of pointlike subsources each emitting independent radiation of the same wavelength but with randomly distributed phase. 
As a consequence of the superposition of the fields of the many subsources a Gaussian field distribution is obtained, resulting in a Boltzmann law for the intensity distribution 
	\begin{equation}
	p(I) = \frac{1}{\langle I\rangle} 
		\exp\left( -\frac{I}{\langle I\rangle} \right),
		\label{eq:intensity_distr}
	\end{equation}
%
%
where $\langle .\rangle$ indicates a time average over time intervals $\Delta t \gg \tau_c$. Here, $\tau_c$  is the characteristic coherence time of the TLS governed by the dynamics of 
the phase fluctuations of the different subsources. 
Note that for a random light source as considered here the intensity average is identical to the sample average since the system is ergodic.\\

		\begin{figure}[h]
\resizebox{\columnwidth}{!}{
  \includegraphics{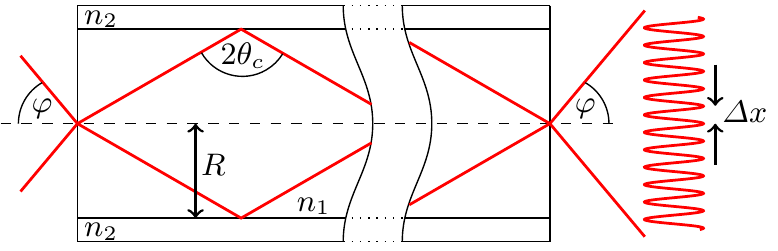}}
\caption{Light propagation in a multimode fiber in a ray model (for details see text).
					\label{fig:fiber_model}}
\end{figure}
To describe our fiber pseudo-TLS we model the propagation of a monochromatic beam in an optical multimode fiber in a ray picture (Fig.\ref{fig:fiber_model}). 
The rays propagate in the fiber due to total internal reflection at the interface fiber core/fiber cladding with refractive index ratio $n_1/n_2 > 1$. Total internal reflection occurs up to the critical angle $\theta_c = \arcsin(n_1 / n_2)$. 
Consequently, the maximal angle under which light can be coupled into or exit the fiber is given by
	\begin{align}
\varphi = \arcsin(\sqrt{n_1^2-n_2^2})=\arcsin(N_A),
\label{number_NA}
	\end{align}
with $N_A$ the numerical aperture of the fiber.

If coherent light is coupled into the fiber a speckle pattern is observed in the far field  behind the fiber 
due to interference between the many modes sustained by the fiber and superposed behind the fiber exit.
Since the optical path length of each mode is determined by the actual geometry of the fiber 
the light in each mode upon propagation through the fiber acquires a different and random phase with respect to the other modes. The modes of the fiber thus imitate the behavior of the many subsources of a TLS.
The smallest size of a subsource at the fiber outlet is determined by the interference pattern produced by the two rays contrapropagating within the fiber at the critical angle \cite{Imai1986}, with a fringe spacing corresponding to (see Fig.~\ref{fig:fiber_model}) 
	\begin{align}
	\Delta x \approx \frac{\lambda}{2} N_A^{-1}.
	\end{align}
With $\Delta x$ one can estimate the number of subsources of a given mode at the fiber outlet 
	\begin{align}
	M_s = \frac{R^2}{\Delta x^2} = \left(\frac{2R\cdot N_A}{\lambda}\right)^2 \; ,
	\label{number_speckle}
	\end{align}
with $2R$ being the diameter of the fiber core.
The total number of modes in the fiber is approximately given by \cite{Gloge1971}
	\begin{align}
	M_m = 2 \left(\frac{\pi R\cdot N_A}{\lambda}\right)^2.
	\label{number_modes}
	\end{align}
By looking at Eqs. (\ref{number_speckle}) and (\ref{number_modes}) one can see that both quantities are of similar magnitude, i.e., on the order of $10^4$ for the fibers used in our experiments.
Moreover, there are about $\binom{10^4}{2}$ mode pairs which can interfere leading to a total number of approximately $\binom{10^4}{2}\cdot 10^4 \approx 10^{11}$ subsources. The phases of the subsources are determined by the $10^4$ random phases of the contributing modes leading in total to a Gaussian field distribution behind the fiber and the observed speckle pattern 
\cite{Imai1986,Takai1985,Doya2002}.

To derive the corresponding photon statistics we measured the intensity distribution $p_F(I)$ in the far field behind the fiber.
To this aim, a linearly polarized HeNe-laser at $633$ nm beam was coupled 
into a standard step index multimode fiber with a core diameter of $200\, \mu$m  and $N_A = 0.39$.
At a distance $z=20\,$cm behind the fiber a monochrome CMOS camera with 8-bit pixel depth  was placed to record a large number of images with different speckle patterns. Since  the gray values of each pixel of the camera are proportional to the impinging intensity one can obtain $p_F(I)$ by simply producing a normalized histogram of the measured pixel intensities for all images.
Note that the linear polarization of the laser light is not maintained during propagation in the multimode fiber. Therefore, a linear polarizer was placed in front of the CMOS camera to measure $p_F(I)$ -- without polarizer one would measure a different distribution, depending on the degree of polarization of the speckle field \cite{Mandel_Wolf1995}.

\begin{figure}[h]
	\resizebox{\columnwidth}{!}{
  \includegraphics{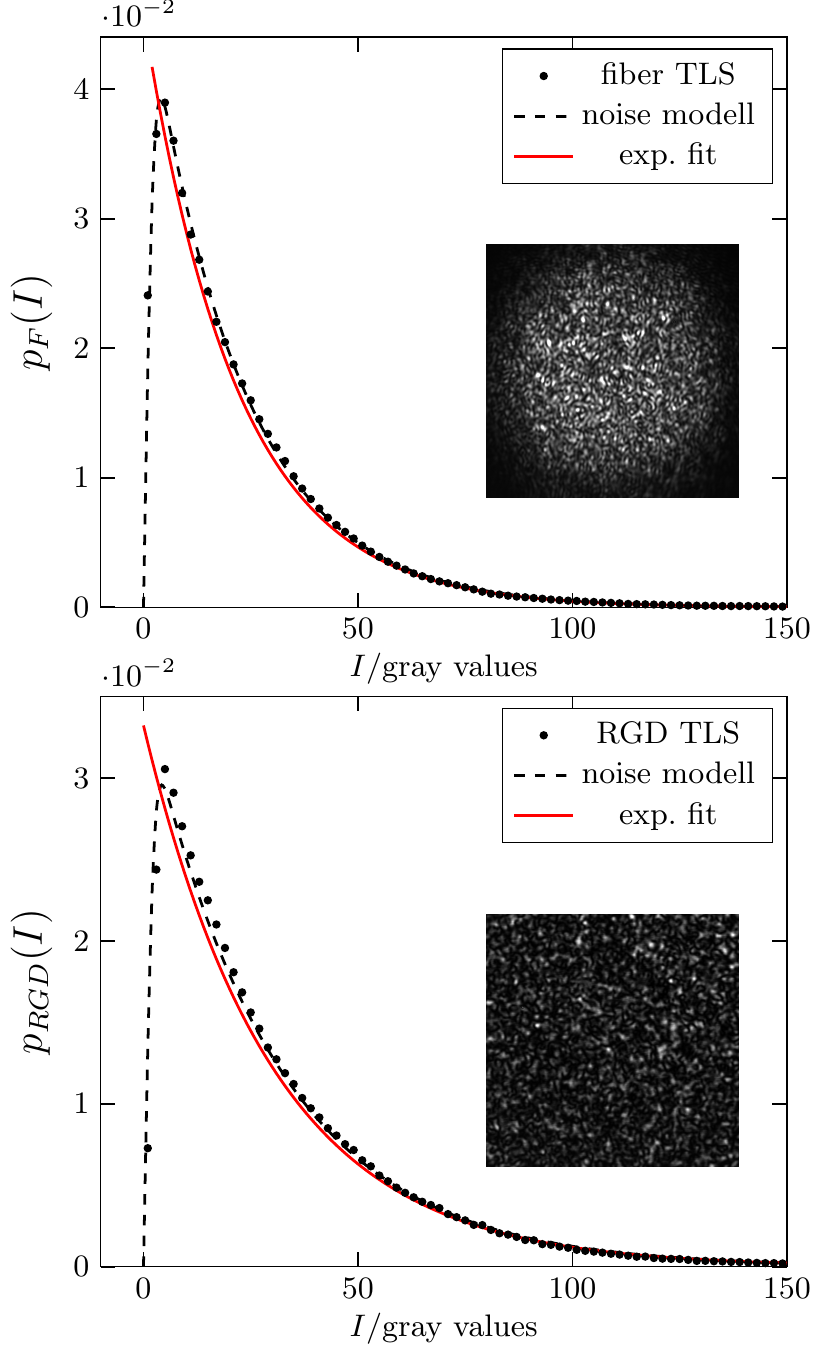}}
\caption{ Normalized intensity distribution of a fiber pseudo-TLS $p_F(I)$  (top) and of a RGD pseudo-TLS $p_{RGD}(I)$  (bottom) measured by a CMOS camera at a distance $z=20\,$cm.
The dashed (black) curves represent the theoretically  expected distributions when incorporating the thermal noise of the CMOS camera to $p(I)$ of Eq.~(\ref{eq:intensity_distr}). The distributions obtained after deconvolving the noise of the camera from the measured values are displayed by the solid (red) curves. They are in excellent agreement with the exponential behaviour of $p(I)$.
The inlays show typical images of the  speckle patterns for the fiber pseudo-TLS (top) and the RGD pseudo-TLS (bottom) as measured by the CMOS camera.
}
	\label{fig:comp_rgd_fiber}
\end{figure}

Fig.~\ref{fig:comp_rgd_fiber} shows the result obtained for $p_F(I)$ under optimal conditions. 
Additionally the distribution $p_{RGD}(I)$ derived from the speckle pattern of a RGD pseudo-TLS is plotted.
As the production of pseudothermal light by use of a RGD is well-known, a comparison of $p_F(I)$ to $p_{RGD}(I)$ is useful to validate the outcome for our fiber pseudo-TLS. In both cases a small deviation from the perfect thermal distribution of Eq.~(\ref{eq:intensity_distr}) can be seen as the measured most probable value is slightly different from zero.
The deviation can be explained by the thermal noise of the CMOS camera. Incorporating this noise to the expected thermal distributions of Eq.~(\ref{eq:intensity_distr}) leads to the black (dashed) curves displayed in Fig.~\ref{fig:comp_rgd_fiber}. The curves are in excellent agreement with $p_F(I)$ and $p_{RGD}(I)$, 
respectively. The corresponding pure thermal statistics - obtained after deconvolving the noise of the camera from the black (dashed) curves - is also shown in Fig.~\ref{fig:comp_rgd_fiber} (see red (solid) curves). The latter are in excellent agreement with the exponential behaviour of $p(I)$ displayed by Eq.~(\ref{eq:intensity_distr}).

Our investigations show that the fiber length and the physical deformation of the fiber do not significantly influence the final result.
The experiments unveal further that it is important to use the entire numerical aperture $N_A$ for the laser-fiber coupling as only in this case the maximum amount of modes are excited in the fiber. 
We achieved this by increasing the laser beam diameter before coupling the laser into the fiber by use of a microscope objective. 
Further, it was found that the maximum of modes are excited not by optimizing the transmitted laser power but 
rather with a slight misalignment of the laser beam with respect to the fiber axis (and thus less transmitted power). In this case an intensity distribution most close to the thermal distribution $p(I)$ of Eq.~(\ref{eq:intensity_distr}) is obtained.

\section{Spatial correlation functions for one and two fiber pseudo-TLS}
\label{sec:corrfunctions}
Next, we investigated the equal-time two-point intensity correlations for one and two independent fiber pseudo-TLS. The visibility of these functions is a further testbed for the gaussianity and the pseudothermal character of the fields generated by the fiber pseudo-TLS. 

The equal-time second order intensity correlation function is given by \cite{Glauber1963} 
\begin{align}
		G^{(2)}(\mathbf{r}_1, \mathbf{r}_2) &= \langle \hat{E}^{(-)}(\mathbf{r}_1) \hat{E}^{(-)}(\mathbf{r}_2)
			\hat{E}^{(+)}(\mathbf{r}_2) \hat{E}^{(+)}(\mathbf{r}_1) \rangle \; ,
		\label{eq:G2}
	\end{align}
where the positive and negative frequency parts of the electric field operator, $\hat{E}^{(+)}(\vec{r})$ and $\hat{E}^{(-)}(\vec{r})$, respectively, take in the far field the form 
\begin{equation}
\left[ \hat{E}^{(-)}(\vec{r}) \right]^{\dagger}	= \hat{E}^{(+)}(\vec{r}) =\sum_{l=1}^{2} e^{i k \vec{n}\cdot \vec{R}_{l}} \hat{a}_{l} \, .
\label{eq:Eplus}
\end{equation}
In Eq.~(\ref{eq:Eplus}), $\hat{a}_{l}$ defines the annihilation operator of a photon emitted from source $l$ at $\vec{R}_{l}$, $k=2 \pi/\lambda$ and $\vec{n}=\vec{r}/|\vec{r}|$ is the direction of propagation of a photon recorded at position $\vec{r}$.

Applying the Gaussian moment theorem \cite{Mandel_Wolf1995} we can simplify Eq. (\ref{eq:G2}) and
write for the normalized equal-time second order intensity correlation
	\begin{align}
		g^{(2)}(\mathbf{r}_1, \mathbf{r}_2) = 1 + \Huge|  \gamma_c (\mathbf{r}_1,\mathbf{r}_2) \Huge|^2 \label{eq:g2} \; ,
	\end{align}
where $\gamma_c (\mathbf{r}_1,\mathbf{r}_2)$ is the complex degree of coherence \cite{Loudon2000}
\begin{align}
	\gamma_c (\mathbf{r}_1,\mathbf{r}_2) = \frac{\langle \hat{E}^{(-)}(\mathbf{r}_1) \hat{E}^{(+)}(\mathbf{r}_2) \rangle}{\langle \hat{E}^{(-)}(\mathbf{r}_1) \hat{E}^{(+)}(\mathbf{r}_1) \rangle \langle \hat{E}^{(-)}(\mathbf{r}_2) \hat{E}^{(+)}(\mathbf{r}_2) \rangle}
\end{align}
According to the Van Cittert-Zernike theorem, the complex degree of coherence of an extended TLS is given by the two-dimensional Fourier transform of its spatial intensity profile $I(\mathbf{r'})$ \cite{Mandel_Wolf1995}, i.e.,
	\begin{align}
	\gamma_c(\mathbf{r}_1, \mathbf{r}_2) &= \frac{\mathcal{F}[I(\mathbf{r'})](\mathbf{\zeta})}
		{\int I(\mathbf{r'})d\mathbf{r'}} \label{vcz_theorem} \\
		\text{where}\quad \zeta
				&=\left( \frac{(x_2-x_1)}{\lambda z}, \frac{(y_2-y_1)}{\lambda z}  \right).
	\end{align}
Here $\mathcal{F}[g(\mathbf{r'})](\mathbf{\zeta})$ denotes the two-dimensional Fourier transform of $g(\mathbf{r'})$ which relates  the source plane $\mathbf{r'}=(x',y')$ to the detector plane $\mathbf{r}=(x,y)$ at a distance $z$.

Considering the end facets of the multimode fibers as extended pseudo-TLS with radius $a$ and constant mean intensity leads to an intensity profile of the  form
	\begin{align}
	&I_1(\mathbf{r'}) = I_0\cdot\text{circ}\left(\frac{|\mathbf{r}'|}{a}\right) & &\text{with}& &\text{circ}(x)
	= \begin{cases}
  1  \,\, \text{for}\,\,  |x| \leq 1 \\
  0  \,\, \text{otherwise.}
	\end{cases} 
	\end{align}
Applying the van Cittert-Zernike theorem yields
	\begin{align}
	\gamma_c(\mathbf{r}_1, \mathbf{r}_2) = \frac{2J_1(\chi)}{\chi} \quad \text{with} \quad 
		\chi = \frac{\pi a}{\lambda z}|\mathbf{r}_1-\mathbf{r}_2| \; ,
			\label{gamma_1TLS}
	\end{align}
where $J_1$ denotes the Bessel function of the first kind.
Combining Eq. (\ref{gamma_1TLS}) with Eq. (\ref{eq:g2}) allows to derive an analytic expression for $g^{(2)}(\mathbf{r}_1, \mathbf{r}_2)$. In the case of one fiber pseudo-TLS one obtains
	\begin{align}
	g_{1\, TLS}^{(2)}(\mathbf{r}_1, \mathbf{r}_2) = 1+
		\left| \frac{2J_1\Big(  \frac{\pi a}{\lambda z} |\mathbf{r}_1-\mathbf{r}_2| \Big)}
				{ \frac{\pi a}{\lambda z} |\mathbf{r}_1-\mathbf{r}_2| } \right|^2.
				\label{eq:g2_1TLS}
	\end{align}
	
In the case of two independent fiber pseudo-TLS, separated by a distance $d$, the intensity profile is given by 
	\begin{align}
	I_2(\mathbf{r'}) = I_0 \left[ \delta\left(x - \frac{d}{2}\right) + 
\delta\left(x + \frac{d}{2}\right) \right]*\text{circ}\left(\frac{|\mathbf{r}'|}{a}\right),
	\end{align}
where  $"*"$ denotes a convolution. Applying the convolution theorem and using Eqs. (\ref{eq:g2}), (\ref{vcz_theorem}) and (\ref{eq:g2_1TLS}) we find 
	\begin{align}
	 g_{2\, TLS}^{(2)}&(\mathbf{r}_1,\mathbf{r}_2) =  \notag \\
		&1 + 
	 \left| \cos\Big(\frac{\pi d}{\lambda z}|\mathbf{r}_1-\mathbf{r}_2| \Big) 
	\cdot \frac{2J_1\Big( \frac{\pi a}{\lambda z} |\mathbf{r}_1-\mathbf{r}_2|\Big)}
			{ \frac{\pi a}{\lambda z} |\mathbf{r}_1-\mathbf{r}_2|} \right|^2.
			\label{eq:g2_2TLS}
	\end{align}
The expression displays an envelope function equal to the one obtained for a single fiber pseudo-TLS, with an additional sinusoidal modulation whose frequency depends on the lateral distance $d$ between the sources.

%
\begin{figure*}[t]
	\resizebox{\textwidth}{!}{
  \includegraphics{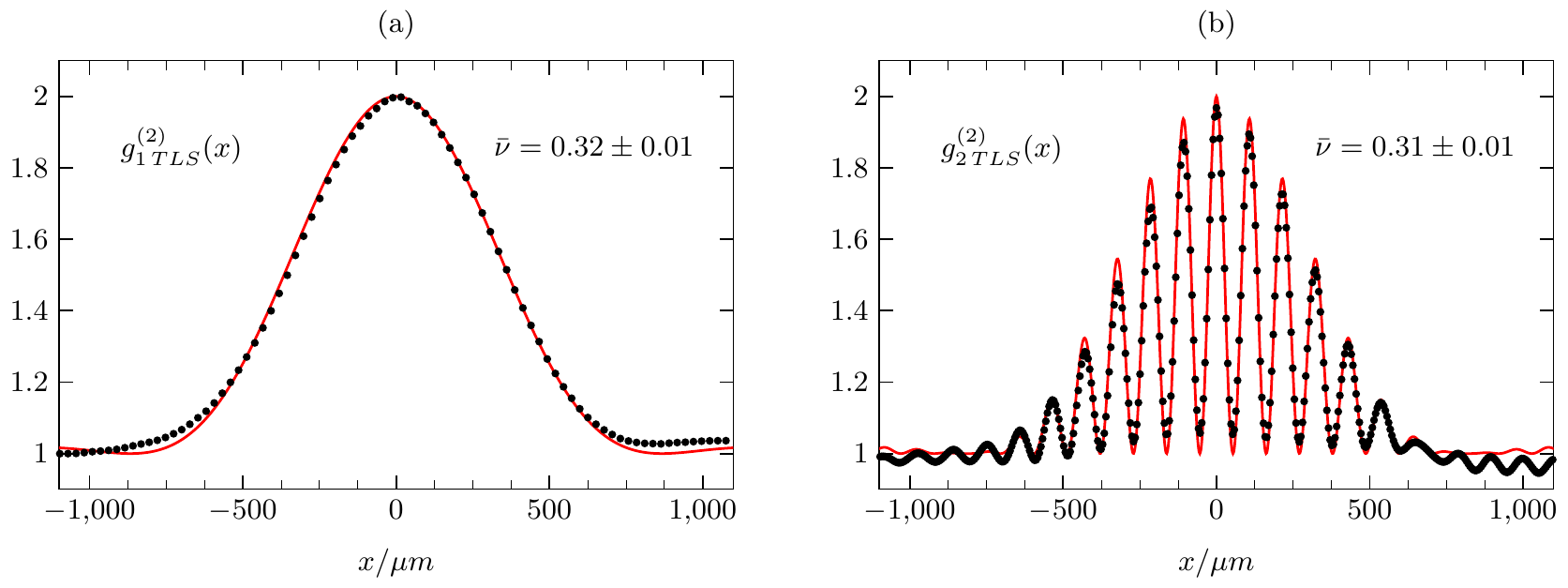}}
\caption{Second order intensity correlations as a function of the relative distance $x=|x_0-x_i|$ between a fixed detector pixel at $x_0$ and a moving detector pixel at $x_i$ for 1 TLS (a) and 2 TLS separated by a distance $d=1\,$mm (b). 
The solid lines represent fits according to Eqs. (\ref{eq:g2_1TLS}) and (\ref{eq:g2_2TLS}), respectively. From a set of $10$ measurements a mean visibility $\bar{\nu} = 0.32 \pm 0.01$ for 1 TLS and $\bar{\nu} =  0.31 \pm 0.01$ for 2 TLS is deduced.
 Both measurements were taken at a distance $z=20\, $cm behind the fiber pseudo-TLS.
}
	\label{fig:meas_results}
\end{figure*}
\begin{figure}[htp]
	\centering{
\resizebox{\columnwidth}{!}{
  \includegraphics{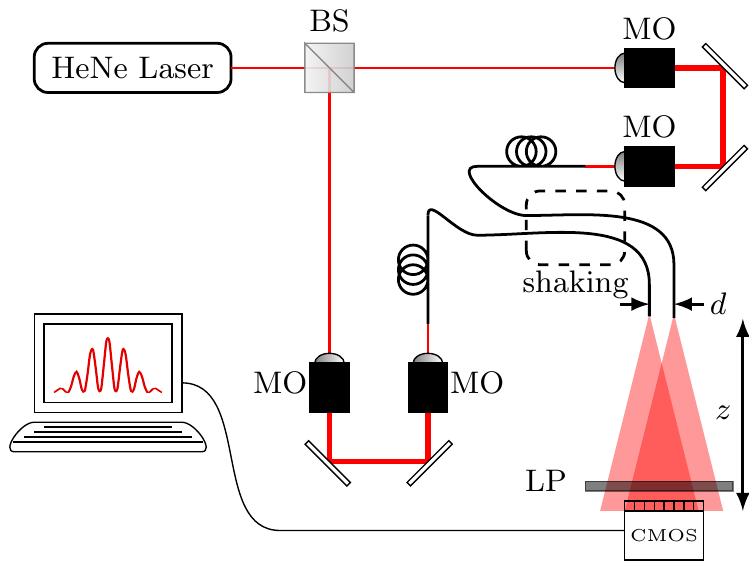}}
\caption{ Experimental setup: monochromatic light of a HeNe-laser is split by a beamsplitter (BS) and coupled into two multimode fibers by use of microscope objevtives (MO) in order to generate two 	independent pseudo-TLS. The spatial intensity correlations are measured by a CMOS camera behind a linear polarizer (LP) at a distance $z$ (for details see text). }
	\label{fig:exp_setup}
	}
\end{figure}
The experimental setup for measuring $g^{(2)}$ 
for one or two fiber pseudo-TLS is shown in Fig. \ref{fig:exp_setup}. As described in Sect. \ref{sec:fiber_as_TLS} light of a linear polarized HeNe laser is coupled into one or both multimode fibers such that each fiber outlet represents an independent fiber pseudo-TLS. 
A physical deformation or movement of the fibers causes the optical path of each fiber mode to vary. Hence, constantly shaking the fibers alters the field composition in the far field behind the fibers, i.e., a different realization of the speckle pattern is continuously obtained, 
just as if rotating the ground glass of a RGD pseudo-TLS.
Due to the complex and discontinuous behavior of the field distribution when shaking the fiber, the temporal behaviour of the fiber pseudo-TLS was not investigated any further. However, for our setup it was found that the speckle pattern changes within a few milliseconds. 

In order to record a large amount of different speckle patterns the CMOS camera was programmed to take images at time intervals $\Delta t \gg 10$ ms. By contrast, the measurement time of each single image was chosen to occur in a time interval $\Delta t_{i} \ll 1$ ms to collect quasi-stationary speckle patterns and thus achieve best visibility. With gray values $I(x_i)$ at each pixel position $x_i$ one-dimensional intensity correlations were computed by correlating $I(x_i)$ along one row of the CMOS chip with $I(x_0)$, where the latter is the gray value at the fixed position $x_0$
	\begin{align}
	g^{(2)}(x_0,x_i) = \frac{\langle I(x_0) I(x_i) \rangle}
													{\langle I(x_0)\rangle \cdot \langle I(x_i) \rangle}.
	\end{align}
A sample average was then carried out over all rows of all images taken by the CMOS camera. 

Fig. \ref{fig:meas_results} shows the corresponding experimental results together with the fitting curves based on Eqs. (\ref{eq:g2_1TLS}) and (\ref{eq:g2_2TLS}), with only the fiber radius $a$, the lateral fiber distance $d$, an additional offset and the visibility as free parameters. As can be seen, the measured  $g^{(2)}$-functions are in excellent agreement with the theoretical expectations. In particular, a visibility of $0.32\pm 0.01 $ and $0.31\pm 0.01$ is obtained for one and two fiber pseudo-TLS, respectively, as expected for a TLS. The results corroborate the assumption that a multimode fiber can indeed serve as a pseudo-TLS. Note that the statistical errors of the data points are below $1\%$ and are thus smaller than the dots displayed in Fig. \ref{fig:meas_results}. The signal-to-noise ratio in both measurements is $S/N > 60$, due to the inherently high directionality of the emitted light and corresponding high photon flux.   

\section{Conclusion and outlook}
 \label{sec:outlook}

In this paper we presented a new pseudo-TLS based on laser light coupled into an optical multimode fiber. The thermal characteristics of the source was verified by measuring the photon statistics as well as the spatial two point intensity correlations for one and two fiber pseudo-TLS. The photons statistics was shown to follow a Boltzmann law to a very high degree whereas the two point intensity correlations displayed a visibility of $0.32 \pm 0.01$ and $0.31 \pm 0.01$ for one and two fiber pseudo-TLS, respectively, close to the theoretically predicted value of $0.33$ for a TLS.
 The experimental outcomes confirm that the fiber pseudo-TLS discussed in this paper fulfill the criteria of a gaussian light source.

Fiber pseudo-TLS have several advantages with respect to other pseudo-TLS. First, they are simple and of low cost using merely a laser system in combination with a short multimode fiber. Further, the light exits the pseudo-TLS within the small solid angle given by the numerical aperture of the fiber. This leads to a high directionality of the emitted light and thus large intensities in the detection plane especially if compared to other pseudo-TLS, e.g., a RGD pseudo-TLS which scatters light in almost all of the half space. This is of particular relevance for experiments working in the low photon counting regime. Finally, the setup displays a high flexibility with respect to the positioning of the sources, e.g., if higher order intensity correlations of several pseudo-TLS are investigated \cite{Oppel2012,Oppel2014,Classen2016}. As shown in \cite{Classen2016}, one can use many fibers and position them rapidly anew in any desired configuration. By contrast, in \cite{Oppel2012,Oppel2014} the $N$ pseudo-TLS were implemented by use of a RGD pseudo-TLS followed by an $N$-slit mask. This setup is static, i.e., for every new arrangement of sources one has to produce and align a novel slit mask.

Potential applications of fiber pseudo-TLS are, e.g., the implementation of Franson interferometers \cite{Franson1989} or the simulation of a controlled-NOT gate \cite{Tamma2016a,Tamma2016b}. Fiber pseudo-TLS might also prove useful for tasks in quantum imaging or quantum metrology, e.g., measurements of higher order spatial intensity correlations in irregular arrangements of pseudo-TLS in 2D. 
\section{Acknowledgments}
The authors gratefully acknowledge funding by the Universit\"atsbund Erlangen-N\"urnberg e.V. and the Erlangen
Graduate  School  in  Advanced  Optical  Technologies (SAOT) by the German Research Foundation (DFG) in the framework of the German excellence initiative.


%

%
%

\end{document}